\documentclass[10pt,twocolumn]{article} 
\usepackage{simpleConference}

\usepackage{graphicx}
\usepackage{newtxtext}
\usepackage{newtxmath}
\usepackage{natbib}
\usepackage{comment}
\usepackage{hyperref}

\newcommand{\ddt}[0]{\frac{d}{dt}}

\usepackage{cancel}

\hypersetup{
    colorlinks = true,
    urlcolor   = blue,
    citecolor  = black,
}

\newcommand{\RomanNumeralCaps}[1]
\linenumbers

\title{Exact expressions for available potential energy and available potential vorticity}

\author{Jeffrey J. Early, Northwest Research Associates, USA \\ Gerardo Hernández-Dueñas, Universidad Nacional Autónoma de México  \\  Leslie M. Smith,University of Wisconsin–Madison, USA \\ M.-Pascale Lelong, Northwest Research Associates, USA}

\begin{document}
\maketitle
\abstract{Exact analytical expressions for available potential energy density (APE) and available potential vorticity (APV) are derived from first principles. These APE and APV expressions align with previously known quantities found using perturbation expansions in \cite{holliday1981-jfm} and \cite{wagner2015-jfm}, respectively. 
The key is to recast the equations of motion and their conservation laws in terms of the coordinate label
$z-\eta$, where $z$ is a fluid parcel's height at the current time, and $\eta$ is the isopycnal deviation from the parcel's current height after adiabatic rearrangement 
of all parcels into the no-motion state. In addition to their intuitive appeal and simplicity, the new APE and APV expressions are easily implemented in numerical computations of Boussinesq dynamics with non-constant stratification.}

% \begin{color}{red}{
% Remaining to-do list, November 22nd, 2022
% \begin{itemize}
%     \item Re-write abstract
%     \item Reference \cite{winters1995-jfm}? [Goal is to separate adiabatic processes from diabatic processes.] Currently in the conclusion. Is this enough?  
%     \item Reference Lorenz?
%     \item Available  potential  energy and the maintenance of the  general  circulation.
%     \item Calculation for $\eta$ in terms of excess density $\rho_\textrm{e}$ 
%     \item Calculation for rigid-lid modes using the PV condition.
%     \item Is $\textrm{APE}_e$ good notation? `exchangeable' APE?
%     \item Should we use $\eta_\textrm{e}$ during the linearization? I think so. 
%     \item check n2 definition
%     % \item Lets assume stratification of a particular (but still general) form, and expand that. Either based on exponential, or log-log. Still general! But provides sensible lowest order expansions.
% \end{itemize}}
% \end{color}

\section{Introduction}
\label{sec:intro}

The conservation laws for energy density\footnote{Energy densities are defined using mass density instead of total mass and hence have dimension of energy per unit volume. We will hereafter refer to `energy density' as simply `energy.'} and potential vorticity (PV) in a rotating Boussinesq fluid are treated in standard geophysical fluid dynamics textbooks \cite[e.g.,][]{vallis2006-book}. However, these quantities are less useful in practice than one might expect because their definitions include a large contribution from the equilibrium state of no motion, which does not play a role in the dynamics of the fluid.  In the case of potential energy, a more useful definition excludes the no-motion contribution, measuring only the potential energy relative to the no-motion state through adiabatic re-arrangement of fluid parcels. Termed available potential energy (APE) by \cite{lorenz1955-tellus}, APE includes a positive definite component available to exchange with kinetic energy (denoted here as $\textrm{APE}_\textrm{e}$) and also allows for a direct assessment of non-adiabatic processes \citep{holliday1981-jfm,winters1995-jfm}. A less obvious consequence of having a nonzero value in the no-motion state is that the small-amplitude limits of the nonlinear PE and PV do not coincide with the PE and PV deduced from the linearized equations of motion. For example, the small amplitude limit of Ertel PV defined with total density does not result in quasi-geostrophic PV (QGPV) for non-constant stratification. As discussed in \cite{early2021-jfm}, only available potential vorticity (APV) as defined in \cite{wagner2015-jfm} has the correct limit.

% So how does one construct APE and APV quantities from first principles? 

The results presented here emerged from our attempt to define equations of motions and conservation laws that include an explicit free-surface which coincide with the conservation laws derived from the linearized equations. However, we found that while the derivations for APE in \cite{holliday1981-jfm} and APV in \cite{wagner2015-jfm} are certainly correct, a more general methodology is still needed. In the results that follow, we present an approach that leads directly to analytical expressions for these quantities that can also be generalized to include additional assumption beyond a rigid upper and lower boundary.

To construct APE and APV we consider the density $\rho_\textrm{tot}({\bf{x}},t)$ of a fluid at a given time and adiabatically rearrange the fluid parcels into their stable no-motion state $\rho_\textrm{nm}(s)$, where 
$\rho_\textrm{nm}$ is a monotonic function of its argument $s$, and where each parcel has a no-motion height $s({\bf x},t)$. For each fluid parcel at the fixed time $t$, the isopycnal deviation $\eta({\bf x},t)$ is the parcel's current height $z$ above its no-motion height $s({\bf{x}},t)$, such that 
$s({\bf x},t) = z - \eta({\bf x},t)$.  
%Then using the no-motion density function $\rho_\textrm{nm}(s)$, 
Our mapping implies the relation $\rho_\textrm{tot}({\bf x},t) = 
\rho_\textrm{nm}(s({\bf x},t)) = \rho_\textrm{nm}(z - \eta({\bf x},t))$.  With this change of variables and using the chain rule, the thermodynamic equation
$\ddt\rho_\textrm{tot}({\bf x},t) = 0$ may be replaced by $\ddt \left(z-\eta({\bf{x}},t)\right)=0$, where $\ddt$ is the total derivative following fluid parcels.  Formulation of the 
potential energy and potential vorticity in terms of the materially conserved label coordinate $z-\eta({\bf x},t)$ is the key to our construction of APE and APV.

\begin{comment}
\vskip 0.1in
\begin{color}{red}{Should we change $s$ to $\tilde{x}$  to match the schematic, or change $\tilde{x}$ to $s$ in the schematic?}
\end{color}
\end{comment}

To arrive at the definition for available potential energy, we use the thermodynamic equation to first construct a new quantity called the materially conserved potential energy (MCPE) which labels each particle with its no-motion potential energy, given by $\textrm{MCPE}=g (z-\eta)\rho_\textrm{nm}(z-\eta)$. Using the properties of material conservation, we show that this leads to the definition of available potential energy
\begin{equation}
\textrm{APE} \equiv g \eta \rho_\textrm{nm}(z-\eta)
\label{def:APE-intro}
\end{equation}
\noindent
which vanishes in the state of no-motion and produces the correct energy exchange term. When using the equations of motion with the hydrostatic background removed, the work done against the hydrostatic pressure to move a fluid parcel from its no-motion position with height $(z-\eta)$ to its current position with height $z$ must be removed from (\ref{def:APE-intro}). This leads to the definition of available potential energy found using a perturbation expansion in
\cite{holliday1981-jfm}.
%The quantity (\ref{def:APE-intro})
%Furthermore, (\ref{def:APE-intro}) exactly closes the energy conservation equation found using a perturbation expansion in 
%\cite{holliday1981-jfm}.
%\begin{color}{red}{Can we be more clear here?} [jje: by balancing the buoyancy flux term that arises in the KE equation?  Leslie needs more discussion...]
%\end{color}

The available potential vorticity is here defined as
\begin{equation}
\textrm{APV} \equiv (\nabla \times {\bf u} + f_0\hat{\bf z}) \cdot \nabla (z-\eta) - f_0
\label{def:APV-intro}
\end{equation}
\noindent
where ${\bf u}$ is the velocity, $f_0$ is the (constant) Coriolis parameter for flow in a frame rotating about the $\hat{\bf z}$-axis.  The quantity \eqref{def:APV-intro}
is conserved following fluid parcels in a rotating Boussinesq fluid and reduces to the quasi-geostrophic potential vorticity when the isopycnal displacement $\eta$ is assumed to be both small and non-overturning.
The definition of potential vorticity (\ref{def:APV-intro}) has been previously noted by \cite{muller1995-rg} and is closely related to the quantity constructed by \cite{wagner2015-jfm} through perturbation expansion.

\section{Background}
\label{sec:background}
%
%%%%%%%%%%%%%%%%%%%%%%%%

Under the Boussinesq approximation, we consider the inviscid equations of motion
\begin{subequations}
\begin{align}
\label{x-momentum}
\partial_t u - f_0 v + \mathbf{u} \cdot \nabla u  =& - \frac{1}{\rho_0} \partial_x p_\textrm{tot}\\ \label{y-momentum}
\partial_t v + f_0 u + \mathbf{u} \cdot \nabla v =&  - \frac{1}{\rho_0}\partial_y p_\textrm{tot}  \\ \label{z-momentum}
\partial_t w  + \mathbf{u} \cdot \nabla w =& - \frac{1}{\rho_0}\partial_z p_\textrm{tot} -\frac{1}{\rho_0}g \rho_\textrm{tot} \\ \label{thermodynamic}
\partial_t \rho_\textrm{tot}  + 
\mathbf{u} \cdot \nabla
\rho_\textrm{tot}=& 0  \\ 
\label{continuity}
\partial_x u + \partial_y v + \partial_z w =& 0 
\end{align}
\label{eqn:boussinesq}
\end{subequations}
where ${\bf u} = (u,v,w)$ is the fluid velocity in Cartesian coordinates, $p_\textrm{tot}$ is the total pressure, $\rho_\textrm{tot}$ is the total density, and $f_0$ is the constant Coriolis parameter. The boundaries are assumed to be periodic in $x$, $y$ and rigid, free-slip boundaries at $z=0$ and $z=-D$. The total pressure and total density contain contributions from the no-motion state when the fluid is described by hydrostatic balance, together with a deviation from the no-motion state. 
Note that there is no mixing allowed since equations \eqref{eqn:boussinesq} do not include viscosity or diffusivity.  Henceforth, we will use the notation $\ddt= \partial_t + \mathbf{u} \cdot \nabla$ to denote the material derivative.

Equations (\ref{eqn:boussinesq})
globally conserve the total energy E = KE + PE, where KE and PE are the kinetic and potential energies, respectively. The equation for $\textrm{KE}=\rho_0 \mathbf{u}^2/2$ follows from \eqref{x-momentum}-\eqref{z-momentum},
\begin{equation}
\label{ke-full}
    \ddt \textrm{KE} = - \mathbf{u} \cdot \nabla p_\textrm{tot}
    - g w \rho_\textrm{tot},
\end{equation}
while the equation for PE requires computing the material derivative of $\textrm{PE} = \rho_\textrm{tot} g z,$ which after application of \eqref{thermodynamic} produces
\begin{equation}
\label{pe-full}
    \ddt  \textrm{PE} = g w \rho_\textrm{tot}.
\end{equation}
Since the buoyancy flux term 
$w g \rho_\textrm{tot}$ matches the equivalent term in the kinetic energy equation \eqref{ke-full}, the sum of the two equations \eqref{ke-full} and \eqref{pe-full} results in the total energy conservation law
\begin{comment}
\begin{equation}
\label{eqn:energy-flux-total}
    \frac{\partial}{\partial t} \left( \textrm{KE} + \textrm{PE} \right) + \mathbf{u} \cdot \nabla \left(  \textrm{KE} + \textrm{PE}  + p_\textrm{tot}  \right) = 0.
\end{equation}
\end{comment}
\begin{equation}
\label{eqn:energy-flux-total}
    \ddt  E= -\mathbf{u} \cdot \nabla p_\textrm{tot}.
\end{equation}
It is noteworthy that the total energy is not materially conserved---while its depth integrated total value is constant in time, the energy of each fluid parcel includes pressure work that depends on the fluid parcel's location within the fluid.  
The available potential energy APE defined in Section \ref{subsec:APE} will have the same energy exchange term as in \eqref{pe-full} and satisfy \eqref{eqn:energy-flux-total}, but will additionally vanish in the no-motion state. 

Another important quantity characterizing (\ref{eqn:boussinesq}) is the potential vorticity 
$\Pi(\psi)$ given by 
\begin{equation}
\label{eqn:pv-background-2}
    \Pi(\psi) \equiv (\nabla \times \mathbf{u} + f_0\hat{z}) \cdot \nabla \psi,
\end{equation}
where $(\nabla \times \mathbf{u} + f_0\hat{z})$ is the total vorticity and  $\psi({\bf x},t)$ is any (scalar) material invariant.  Under the Boussinesq approximation \eqref{eqn:boussinesq} and if $\psi = \psi(\rho_\textrm{tot})$, then the potential vorticity $\Pi$ itself is also a material invariant such that
\begin{equation}
\label{eqn:Piconservatioin}
    \ddt \Pi = 0.
\end{equation}
Traditionally, $\psi$ is chosen 
as $\rho_\textrm{tot}({\bf x},t)$ itself, but this choice results in a definition of PV that does not vanish with no-motion, has a large Eulerian signature for internal waves, and linearizes to a definition inconsistent with the linearized equations of motion. In section \ref{subsec:APV}, 
we will discuss merits of the choice $\psi({\bf x},t) = z-\eta({\bf x},t) = \rho_\textrm{nm}^{-1}(\rho_\textrm{tot}({\bf x},t))$, where the no-motion density function $\rho_\textrm{nm}(z)$ is defined in the next section \ref{subsec:nomotionsoln}.

\begin{comment}
A general definition of potential vorticity $\Pi$ is given by 
\begin{equation}
\label{eqn:pv-background}
    \Pi(\psi) \equiv (\nabla \times \mathbf{u} + f_0\hat{z}) \cdot \nabla \psi,
\end{equation}
where the function $\psi$ is a scalar quantity that is materially conserved by \eqref{eqn:boussinesq}, such as total density $\rho_\textrm{tot}$.
If $\psi$ satisfies the condition
\begin{equation}
\label{eq:PsiCond}
\nabla \psi \cdot (\nabla p_\textrm{tot} \times \nabla \rho_\textrm{tot}^{-1}) = 0,
\end{equation}
which implies $\psi = \psi(p_\textrm{tot},\rho_\textrm{tot})$, then $\Pi$ is a material invariant such that
\begin{equation}
    \frac{\partial \Pi}{\partial t} + \mathbf{u} \cdot \nabla  \Pi = 0.
\end{equation}
We shall discuss the choice $\psi = z-\eta$ in Section \ref{subsec:APV}.
\end{comment}

%%%%%%%%%%%%%%%%%%%%%%%%
\subsection{The no-motion solution}
\label{subsec:nomotionsoln}
%%%%%%%%%%%%%%%%%%%%%%%%

\begin{figure}
\begin{center}
{\includegraphics[width=0.4 \textwidth]{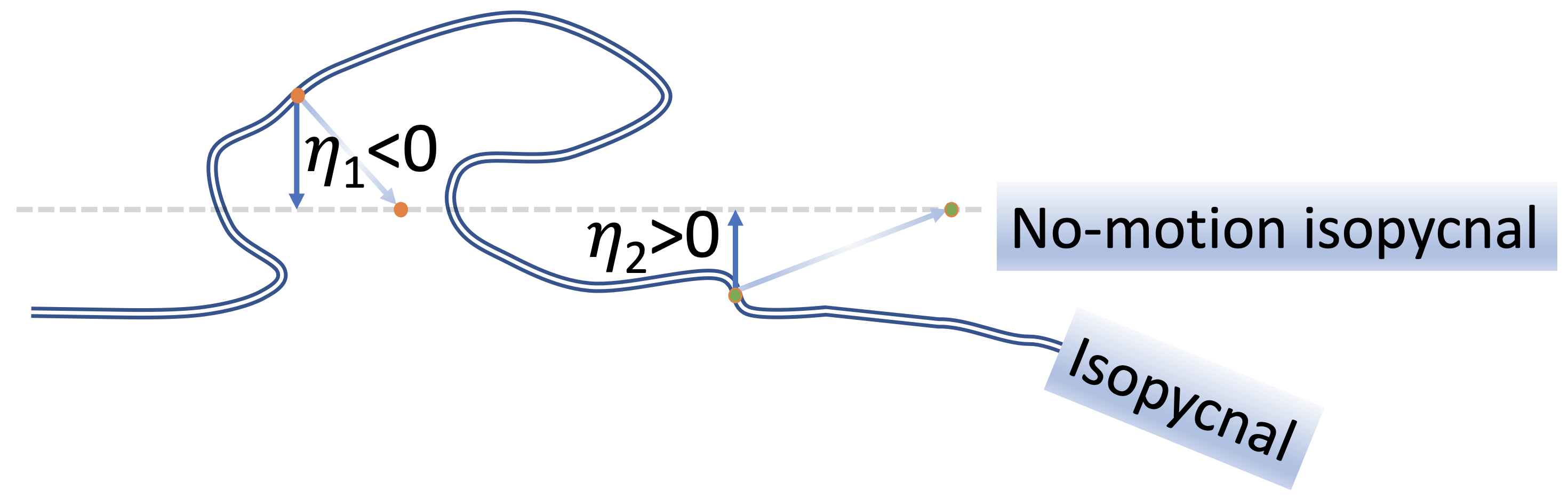}}
\caption{\label{fig:Schematic} Schematic of adiabatic re-arrangement and isopycnal displacements.}
\end{center}
\end{figure}

The most trivial, but important, solution to the equations of motion occurs when we let ${\bf u}({\bf x},t) = (u,v,w)=(0,0,0)$, where the position vector ${\bf x} = (x,y,z)$ and $z$ measures height from the ocean bottom $z = -D$. The density of the fluid $\rho_\textrm{tot}({\bf x},t)$ must be adiabatically re-arranged to eliminate horizontal pressure and density gradients as sketched in Figure \ref{fig:Schematic}. This `no-motion' solution is denoted as $(u,v,w,p,\rho)=\left(0,0,0,p_\textrm{nm}(z),\rho_\textrm{nm}(z) \right)$ where
\begin{equation}
\label{nm-sol}
     p_\textrm{nm}(z) = -g \int_0^z  \rho_\textrm{nm}(\xi) d\xi
\end{equation}
so that $\partial_z p_\textrm{nm}(z) \equiv -g \rho_\textrm{nm}(z)$ and the equations of motion are satisfied. The no-motion, hydrostatic solution (\ref{nm-sol}) may be used to define excess pressure, or perturbation pressure as,
\begin{equation}
\begin{split}
p_\textrm{e}(\mathbf{x},t) \equiv& p_\textrm{tot}(\mathbf{x},t) - p_\textrm{nm}(z) \\ =&  p_\textrm{tot}(\mathbf{x},t) + g \int_0^{z} \rho_\textrm{nm}(\xi) d\xi.
\end{split}
\end{equation}

%%%%%%%%%%%%%%%%%%%%%%%%
\subsection{Isopycnal deviation vs excess density}
\label{subsec:iso-vs-excess}
%%%%%%%%%%%%%%%%%%%%%%%%

There are two common approaches to expressing re-arrangements of density relative to the no-motion solution: isopycnal deviation ($\eta$) or excess density ($\rho_\textrm{e}$) defined by
\begin{subequations}
\label{iso-and-excess-def}
\begin{align}
\label{excess-def}
    \rho_\textrm{tot}({\bf x},t) \equiv& \rho_\textrm{nm}(z) + \rho_\textrm{e}({\bf x},t) \textrm{ or} \\
\label{iso-def}
    \rho_\textrm{tot}({\bf x},t) \equiv& \rho_\textrm{nm}(z - {\eta}({\bf x},t)).
\end{align}
\end{subequations}
Both formulations disallow mixing in order to maintain material conservation, but do allow isopycnal overturning. Note that the argument of $\rho_\textrm{nm}$ in \eqref{iso-def} could be used to define an isopycnal coordinate (lines of constant $z-\eta$), but this would disallow isopycnal overturns. In practice, computing $\eta$ from $\rho_\textrm{e}$ requires inverting $\rho_\textrm{nm}$ in equation \eqref{iso-def} with either an analytically specified no-motion density profile, or performing an adiabatic re-arrangement of an existing density field using, e.g., equation (2.3) in \cite{winter2013-jfm}. For analytical work, it is useful to approximate the relationship between $\rho_\textrm{e}$ and $\eta$. Using $\rho_\textrm{nm}(z)$ and its inverse from the definitions in \eqref{iso-and-excess-def} results in series expansions
\begin{equation}
\begin{split}
 \eta = - \sum_{n=1}^\infty \frac{1}{n!} \left( \frac{1}{ \partial_z \rho_\textrm{nm}(z)} \frac{d}{dz} \right)^{n-1} \left( \frac{1}{\partial_z \rho_\textrm{nm}(z)} \right)  \rho_\textrm{e}^n  \\ \rho_\textrm{e} = 
    \sum_{n=1}^\infty \frac{(-1)^n}{n!} \rho^{(n)}_\textrm{nm}(z) \eta^n
\end{split}
\end{equation}
where the differential operator in the $\eta$ expansion follows from implicit differentiation of $\frac{d}{d \rho_\textrm{nm}}$. The first-order approximation to isopycnal deviation written in terms of excess density is thus given by $\eta_\textrm{e} \equiv - \rho_\textrm{e} \left( \partial_z \rho_\textrm{nm} \right)^{-1}$ and the first terms in the expansion are
\begin{equation}
\begin{split}
    \eta =& \eta_e + \frac{1}{2} \partial_z \log \left( N^2(z) \right) \eta_e^2 \\ &+ \frac{1}{12} N^4 \partial_{zz} \left( N^{-4} \right) \eta_e^3  + O\left( \eta_\textrm{e}^4 \right)
    \end{split}
\end{equation}
where the squared buoyancy frequency is defined as $N^2 = -g \partial_z \rho_\textrm{nm}/\rho_0$.

A key constraint is that the density of the fluid is bounded by $\left[ \rho_0, \rho_D \right]$, where $\rho_0 = \rho_\textrm{nm}(z=0)$ and $\rho_D = \rho_\textrm{nm}(z=-D)$. No fluid parcels exist with a density less than $\rho_0$ or greater than $\rho_D$, which requires
\begin{subequations}
\begin{align}
    &\rho_0 \leq \;  \rho_\textrm{nm}(z) + \rho_\textrm{e}({\bf x},t) \leq \rho_{D} \quad \textrm{or} \\
     &z \leq \;  \eta({\bf x},t) \leq z+D.
\end{align}
\end{subequations}
Material conservation of density further requires 
\begin{subequations}
\begin{align}
\label{excess-volume-conservation}
\int \rho_\textrm{e} dV &=0 \textrm{ or} \\
\int \eta dV =& 0,
\label{iso-volume-conservation}
\end{align}
\end{subequations}
% \begin{equation}
% \nonumber
%   \int \rho_\textrm{e} dV =0 \quad  \textrm{(2.13a)} \quad \textrm{ or } \quad
% \int \eta dV = 0, \quad \textrm{(2.13b)}
% \end{equation}
% \begin{equation}
%   \int \rho_\textrm{e} dV =0  \quad \textrm{ or } \quad
% \int \eta dV = 0
% \end{equation}
where $\int dV$ denotes integration over the entire volume of fluid. These two different approaches affect only the vertical momentum equation \eqref{z-momentum} and the thermodynamic equation \eqref{thermodynamic}. Starting with the vertical momentum equation \eqref{z-momentum} we subtract the no-motion solution \eqref{nm-sol} to express two forms of the vertical momentum equation, 
\begin{subequations}
\begin{align}
\ddt w
=& - \frac{1}{\rho_0} \partial_z p_\textrm{e}- \frac{1}{\rho_0} g \rho_\textrm{e}  \textrm{ or}\\
\ddt w
=& - \frac{1}{\rho_0} \partial_z p_\textrm{e}- \frac{1}{\rho_0} g \left( \rho_\textrm{nm}(z - {\eta})- \rho_\textrm{nm}(z) \right).
\end{align}
\label{eqn:w-momentum-excess}
\end{subequations}
To close the equations of motion, we must also express the thermodynamic equation in terms of these two quantities, e.g., 
\begin{subequations}
\begin{align} \label{thermodynamic-excess}
    \frac{1}{\partial_z \rho_\textrm{nm}} \ddt  \rho_\textrm{e} + w =& \; 0  \textrm{ or} \\ \label{thermodynamic-iso}
    \ddt  {\eta} - w =& \; 0.
\end{align}
\end{subequations}
Any number of variations of these expressions are possible. For example, rewriting the excess density conservation \eqref{thermodynamic-excess} using the definition of $\eta_\textrm{e}$
results in an exact equation for excess density conservation $\ddt  \eta_\textrm{e} - w = - w \eta_\textrm{e} \partial_z \left( \ln N^2 \right)$
that resembles the isopycnal thermodynamic equation \eqref{thermodynamic-iso}, but contains an extra when $N^2$ is not uniform.

%%%%%%%%%%%%%%%%%%%%%%%%
\subsection{Material conservation and interpretation of $z-\eta$}
\label{subsec:mat-cons-of-z-eta}
%%%%%%%%%%%%%%%%%%%%%%%%

The primary advantage to using isopycnal deviation $z-\eta({\bf x},t)$ instead of excess density $\rho_\textrm{e}({\bf x},t)$ is that $z-\eta({\bf x},t)$ is a materially conserved quantity that can be treated like a coordinate. Material conservation of $z-\eta({\bf x},t)$ follows directly from material conservation of $\rho_\textrm{tot}({\bf x},t)$ \eqref{thermodynamic} and the definition of $\eta$ \eqref{iso-def}. Going further, \emph{any} quantity $f$ expressed in terms of $z-\eta({\bf x},t)$ is material conserved,
\begin{equation}
\label{eqn:fpe}
\ddt f(z-\eta) = 0.
\end{equation}
We will use this flexibility to define a materially conserved potential energy below.

The volume integral of a material conserved quantity $f(z-\eta)$ remains constant and thus must match the volume integral of any other configuration of the fluid, including the configuration where $\eta=0$, such that
\begin{equation}
\label{eqn:fpe-volume-int}
    \int f(z-\eta) dV = \int f(z) dV.
\end{equation}
This last expression says nothing about the dynamics of how the fluid moved between the different states of motions, just that fluid labels have to be conserved during such a process.

% \textcolor{red}{I was thinking that maybe the conceptual way to argue this is to think of f like a passive tracer—for a passive tracer the total volume integral will result in the same constant, no matter the configuration of fluid or how it got there (or even whether you changed the energy or PV of the fluid).}

% \textcolor{red}{But, this also has me thinking, our `available' definitions really do \emph{not} mean you can reconfigure the fluid with the same PV or same energy.  }

%%%%%%%%%%%%%%%%%%%%%%%%

\section{Available potential energy and available potential vorticity}
\label{sec:ape-apv}

Any definition of available potential energy and vorticity should both appeal to our intuition, and meet the necessary dynamical requirements such as energy exchange and material conservation. We demand four requirements of our definitions of APE and APV.

We describe the four requirements starting with an intuitive definition of APE. \cite{vallis2006-book} writes `The difference between the total potential energy of the fluid and the total potential energy after an adiabatic rearrangement to a state in which the isentropic surfaces are flat is called the available potential energy, or APE.' Thus, the first requirement is the {\it global requirement} that the volume integral of APE (APV), must be the difference between the volume integrals of the PE (PV) in the current and no-motion states. An even stronger condition is to require that the APE (APV) must vanish point-wise in the limit of no motion, which we refer to as  the {\it no motion requirement.} Third, the material derivative of the APE must produce the energy exchange term $g w \rho_\textrm{nm}(z-\eta)$, while the material derivative of the APV must vanish, or in other words there is {\it a material requirement.}
The forth constraint demands that the linearized versions of APE (APV) must coincide with the expressions derived from the linearized equations of motion, 
{\it i.e., a linearization requirement}.

\begin{comment}
In total, there are four requirements that we demand of our definitions of `available' potential energy and vorticity:
\begin{enumerate}
\item 
\textbf
{No-motion requirement} 
The APE (APV) must vanish point-wise in the limit of no motion.  We call this the {\it no motion requirement.}
    \item \textbf{Material requirement} The material derivative the APE must produce the energy exchange term $g w \rho_\textrm{nm}(z-\eta)$, while the material derivative of the APV must vanish.
    \item \textbf{Global requirement} The volume integral of APE (APV) must be the difference between the volume integral of the PE (PV) of the fluid in its current state, minus the fluid in its state of no motion.
    \item \textbf{Linearization requirement} Equivalent expressions for APE (APV) derived from the linearized equations of motion must coincide with linearized version of APE (APV).
\end{enumerate}
\end{comment}

% The global requirement automatically follows as a consequence of  the no-motion requirement, but we keep it as a separate line item for clarity. The no-motion requirement can also be seen as a necessary step toward meeting the linearization requirement.

In the remainder of this section we explain the physical and mathematical reasoning to arrive at the proposed expressions for APE \eqref{def:APE-intro} and APV \eqref{def:APV-intro} that meet the first three requirements. Section \ref{sec:eom-summary} shows that these definitions also meet the linearization requirement.

\label{sec:APE-APV}

%%%%%%%%%%%%%%%%%%%%%%%%
\subsection{Available potential energy (APE)}
\label{subsec:APE}
%%%%%%%%%%%%%%%%%%%%%%%%

% Consider the kinetic energy equation \eqref{ke-full},
% \begin{equation}
% \label{ke}
%     \frac{D}{Dt} \textrm{KE} = - \mathbf{u} \cdot \nabla p_\textrm{tot}
%     - g w \rho_\textrm{nm}(z - \eta)
% \end{equation}
% where we have used \eqref{iso-def} to describe the current density state as $\rho_\textrm{nm}(z - \eta)$. In the context of \eqref{ke}, the goal in defining an equation for APE is to find an expression with a matching energy exchange term, namely $gw \rho_\textrm{nm}(z-\eta)$. 

% [jje: Note that definition of PE leads to 1) correct exchange term and 2) Vallis definition. While the traditional approach satisfies 1, it cannot accomplish 2, b/c it does not have material conservation.]

The key to constructing a suitable definition for APE that meets all requirements is to define a materially conserved potential energy 
\begin{equation}
\label{eqn:mcpe}
\textrm{MCPE} \equiv g (z-\eta)\rho_\textrm{nm} (z-\eta)\;\; \textrm{such that} \;\; \ddt \textrm{MCPE}=0,
\end{equation}
which follows from statement \eqref{eqn:fpe} and labels each particle with its potential energy in the state of no-motion, $z-\eta$. Expanding \eqref{eqn:mcpe} leads to 
\begin{equation}
\label{eqn:pe-flux}
    \ddt  \left[ g \eta \rho_\textrm{nm}(z - \eta) \right] =  g w \rho_\textrm{nm}(z - \eta),
\end{equation}
where the right-hand-side of \eqref{eqn:pe-flux} is the matching energy exchange term from the kinetic energy equation \eqref{ke-full}. As stated in \eqref{pe-full}, this is the same energy exchange term arising from the 
the material derivative of the full  
$\textrm{PE} = g z \rho_\textrm{tot}$. 
%the traditional approach takes material derivative of $g z \rho_\textrm{tot}$ and recovers the same energy exchange term. 
However, in this case, because we have material conservation of the MCPE, we can apply \eqref{eqn:fpe-volume-int}, 
%and reference the volume integral to the state of no-motion. This leads to the relation
leading to 
\begin{equation}
\label{eqn:ape-volume}
    \int g z \rho_\textrm{nm}(z - \eta) dV - \int g z \rho_\textrm{nm}(z) dV  = \int g \eta \rho_\textrm{nm}(z - \eta) dV,
\end{equation}
which is the mathematical statement of the global requirement for a good definition of APE.\footnote{\cite{vallis2006-book} derives an expression for APE from the global requirement (lhs of \eqref{eqn:ape-volume}), but assumes the fluid is non-overturning by using isopycnal coordinates. This results in a different expression for APE.}   

Based on \eqref{eqn:mcpe}-\eqref{eqn:ape-volume}, we propose the definition
\begin{equation}
\label{eqn:pe-eta}
    \textrm{APE} \equiv g \eta \rho_\textrm{nm}(z - \eta),
\end{equation}
which satisfies the no-motion requirement by definition, since it vanishes for $\eta = 0$.  Furthermore, we have demonstrated that it satisfies the material requirement \eqref{eqn:pe-flux}, and the global requirement \eqref{eqn:ape-volume}.
Using this definition of APE, we can now replace the total energy conservation law \eqref{eqn:energy-flux-total} by a version that uses APE instead of PE
\begin{equation}
\label{eqn:energy-flux-pe}
    \ddt  \left( \textrm{KE} + \textrm{APE} \right)=  - \mathbf{u} \cdot \nabla p_\textrm{tot}.
\end{equation}

%%%%%%%%%%%%%%%%%%%%%%%%
\subsection{Exchangeable available potential energy ($\textrm{APE}_\textrm{e}$)}
\label{subsec:APEe}
%%%%%%%%%%%%%%%%%%%%%%%%

In the original derivation of available potential energy, \cite{holliday1981-jfm} start with a kinetic energy equation using excess pressure $p_\textrm{e}$ instead of total pressure $p_\textrm{tot}$, that is using (\ref{x-momentum})-(\ref{y-momentum}) with (\ref{eqn:w-momentum-excess}) instead of (\ref{x-momentum})-(\ref{z-momentum}).  Following from (\ref{eqn:w-momentum-excess}), this kinetic energy equation can be written in 2 different ways:
\begin{subequations}
\begin{align}
\ddt \textrm{KE} =& - \mathbf{u} \cdot \nabla p_\textrm{e}
    - g w \rho_\textrm{e}(z),\; \textrm{or}\\ \label{eqn:KE-excess-eta}
    \ddt \textrm{KE} =& - \mathbf{u} \cdot \nabla p_\textrm{e}
    - g w \left( \rho_\textrm{nm}(z - {\eta}) -\rho_\textrm{nm}(z) \right).
\end{align}
\label{eqn:KE-excess}
\end{subequations}
Thus the same four requirements for a good definition of excess APE are applicable here, except that the material requirement now has an extra term that appears on the right-hand-side of the kinetic energy equation \eqref{eqn:KE-excess-eta}. To arrive at their definition of excess APE, \cite{holliday1981-jfm} perform a perturbation expansion of the term $\rho_\textrm{nm}(z - {\eta}) -\rho_\textrm{nm}(z)$. To arrive at the same answer, we generalize the MCPE \eqref{eqn:mcpe} to also include the hydrostatic pressure from each particle's no-motion position. That is, we define
\begin{equation}
\label{eqn:mcpe-e}
\textrm{MCPE}_\textrm{e} \equiv g (z-\eta)\rho_\textrm{nm} (z-\eta) + p_\textrm{nm}(z-\eta)
\end{equation}
where the additional term $p_\textrm{nm}(z-\eta)$ is materially conserved and thus the $\textrm{MCPE}_e$ is also materially conserved.
%to also include the hydrostatic pressure from each particle's no-motion position. 
The 
%resulting 
volume integral of \eqref{eqn:mcpe-e} leads to 
%with respect to the no-motion state leads to the definition of $\textrm{APE}_\textrm{e}$,
\begin{equation}
\label{eqn:ape_e-volume}
    \int g z \rho_\textrm{nm}(z - \eta) dV - \int g z \rho_\textrm{nm}(z) dV  = \int \textrm{APE}_\textrm{e} \, dV,
\end{equation}
where
\begin{equation}
\label{eqn:pe-hm}
    \textrm{APE}_\textrm{e} \equiv g \eta \rho_\textrm{nm}(z - \eta) + p_\textrm{nm}(z) - p_\textrm{nm}(z-\eta).
\end{equation}
This definition of $\textrm{APE}_\textrm{e}$ meets all four requirements for a good definition of excess available potential energy. It is worth noting that because $p_\textrm{nm}(z-\eta)$ is a materially conserved quantity, \eqref{eqn:fpe-volume-int} implies that the volume integral of $\textrm{APE}_\textrm{e}$ and APE are identical. The relation \eqref{eqn:pe-hm} is equation (3.1) in \cite{holliday1981-jfm}, equation (2.1) in \cite{roullet2008-jfm}, and equation (3.3) in \cite{winter2013-jfm}.

So how can we understand the differences between $\textrm{APE}_\textrm{e}$ and APE? The key is that $\textrm{APE}_\textrm{e}$ satisfies the excess energy flux equation
\begin{equation}
\label{eqn:energy-flux-pe-hm}
    \ddt  \left( \textrm{KE} + \textrm{APE}_\textrm{e} \right) = -\mathbf{u} \cdot \nabla p_\textrm{e},
\end{equation}
while APE \eqref{eqn:pe-eta} satisfies the total energy flux equation \eqref{eqn:energy-flux-pe}. The difference is that $\textrm{APE}_\textrm{e}$ can be seen as neglecting the work done against hydrostatic pressure to move the particle from its no-motion position at $z-\eta$ to its current position at $z$. To see this, note that the hydrostatic pressure $p_\textrm{nm}(z)$ results in a force given by $\mathbf{F}_p = - \partial_z p_\textrm{nm}(z) \hat{z}$, and thus
\begin{equation}
    \textrm{APE}_\textrm{e} = \textrm{APE} - \int_{z-\eta}^z F_p dz^\prime.
\end{equation}
The work done on a fluid parcel as it moves through the fluid can only be correctly diagnosed using the KE plus APE, which includes work done by all the forces in the fluid. Thus any Lagrangian time series or assessment of power should use the APE. However, if the goal is to assess the potential energy of the fluid available to exchange with kinetic energy, then $\textrm{APE}_\textrm{e}$ is the better metric, as shown with several good examples in \cite{winter2013-jfm}.

\subsection{Available potential vorticity}
\label{subsec:APV}

\begin{comment}
Potential vorticity is defined as,
\begin{equation}
\label{eqn:pv}
    \Pi(\psi) \equiv (\nabla \times \mathbf{u} + f_0\hat{z}) \cdot \nabla \psi,
\end{equation}
where the function $\psi$ can be any materially conserved quantity.  If $\psi$ satisfies the condition
\begin{equation}
\label{eq:PsiCond}
\nabla \psi \cdot (\nabla p_\textrm{tot} \times \nabla \rho_\textrm{tot}^{-1}) = 0,
\end{equation}
which implies $\psi = \psi(p_\textrm{tot},\rho_\textrm{tot})$, then $\Pi$ is also a material invariant such that
\begin{equation}
    \frac{\partial \Pi}{\partial t} + \mathbf{u} \cdot \nabla  \Pi = 0.
\end{equation}
\end{comment}

In \eqref{eqn:pv-background-2}-\eqref{eqn:Piconservatioin}, $\psi$
is traditionally 
taken to be the density, i.e., $\rho_\textrm{nm}(z-\eta)$, which results in a definition of potential vorticity $\Pi(\psi)$ that includes a large non-zero value for each fluid parcel even in the state of no-motion, namely $f_0 \partial_z \rho_\textrm{nm}(z)$. The difficulty here is then defining a materially conserved form of potential vorticity that is zero in the state of no-motion.

On the other hand, choosing
$\psi = z - \eta$, one finds
\begin{equation}
\label{eqn:pv-eta}
\textrm{PV} \equiv 
(\nabla \times {\bf u})\cdot \hat{\bf z} + f_0 - f_0 \partial_z \eta - (\nabla \times {\bf u}) \cdot \nabla \eta,
\end{equation}
exactly as written in \cite{muller1995-rg}, his equation (107), although he erroneously states that overturns are not allowed. The expression for PV \eqref{eqn:pv-eta}  can trivially be converted into an available potential vorticity by removing the constant $f_0$ as in \eqref{def:APV-intro}, and thus we arrive at
\begin{equation}
\label{eqn:apv-eta}
    \textrm{APV} \equiv 
(\nabla \times {\bf u})\cdot \hat{\bf z} - f_0 \partial_z \eta - (\nabla \times {\bf u}) \cdot \nabla \eta.
\end{equation}
Following the principle in \eqref{eqn:fpe-volume-int}, the volume integral of available potential vorticity vanishes, $\int \textrm{APV} dV = 0$, and all four requirements for a good definition of APV are met. As potential vorticity has both positive and negative values that lead to the vanishing of its volume integral, it is useful to define the quadratic enstrophy as $Z\equiv \frac{1}{2} \Pi^2$, which is also materially conserved.

\cite{wagner2015-jfm} define available potential vorticity as `the difference between the total PV and the PV arising by advection of the background buoyancy field'. They use this idea to construct a quantity defined as
\begin{equation}
\label{eqn:apv-wy-def}
    \textrm{APV}_\textrm{wy} \equiv \Pi\left[ \rho_\textrm{nm} + \rho_\textrm{e}\right] - \Pi_\textrm{nm}\left[ \rho_\textrm{nm} + \rho_\textrm{e}\right]
\end{equation}
where $\Pi_\textrm{nm}(z)$ is a function that maps the current density $\rho_\textrm{tot}({\bf x},t) = \rho_\textrm{nm}(z) + \rho_\textrm{e}({\bf x},t)$ back to its no-motion PV state, such that $\Pi_\textrm{nm}\left[ \rho_\textrm{nm} + \rho_\textrm{e}\right] = f_0 \partial_z \rho_\textrm{nm}$. 
Then they assume $\rho_\textrm{e} \ll \rho_\textrm{nm}$ to construct a perturbation series in powers of $\rho_\textrm{e}$. The APV \eqref{eqn:apv-eta} results from using
\begin{equation}
\label{eqn:apv-alt-def}
    \textrm{APV} \equiv \Pi\left[ \rho_\textrm{nm}^{-1} \left(\rho_\textrm{nm} + \rho_\textrm{e} \right)\right] - \Pi_\textrm{nm}\left[ \rho_\textrm{nm}^{-1} \left(\rho_\textrm{nm} + \rho_\textrm{e} \right)\right]
\end{equation}
where $\rho_\textrm{nm}^{-1} \left(\rho_\textrm{nm} + \rho_\textrm{e} \right) = z - \eta$. 
The key advantage to this approach is that it does not require a perturbation expansion because $\Pi_\textrm{nm}\left[ \rho_\textrm{nm}^{-1} \left(\rho_\textrm{nm} + \rho_\textrm{e} \right)\right]=f_0$ and thus \eqref{eqn:apv-alt-def} leads directly back to the definition of APV in \eqref{eqn:apv-eta}.
The net result is that APV \eqref{eqn:apv-eta} is an exact, closed form expression potential vorticity conserved in this system.

%%%%%%%%%%%%%%%%%%%%%%%
\section{Summary of equations of motion}
\label{sec:eom-summary}
%%%%%%%%%%%%%%%%%%%%%%%

Here we recapitulate the nonlinear equations of motion and their associated conservation laws, and then demonstrate how the linearized equations lead to conservation laws consistent with low order expansions of the nonlinear conservation laws. The equations of motion
\begin{subequations}
\begin{align}
\label{x-momentum-eta}
\partial_t u - f_0 v + \mathbf{u} \cdot \nabla u  &= - \frac{1}{\rho_0} \partial_x p_\textrm{tot}\\ \label{y-momentum-eta}
\partial_t v + f_0 u + \mathbf{u} \cdot \nabla v &= -\frac{1}{\rho_0} \partial_y p_\textrm{tot}  \\ \label{z-momentum-eta}
\partial_t w  + \mathbf{u} \cdot \nabla w &= - \frac{1}{\rho_0}  \partial_z p_\textrm{tot} -\frac{1}{\rho_0}g \rho_\textrm{nm}(z-\eta) \\ \label{thermodynamic-eta}
\partial_t \eta  + \mathbf{u} \cdot \nabla \eta  &= w  \\ \label{continuity-eta}
\partial_x u + \partial_y v + \partial_z w &= 0, 
\end{align}
\label{eqn:boussinesq-eta}
\end{subequations}
have energy quantities and potential vorticity
\begin{subequations}
\begin{align}
\label{eqn:ape-apv-recap}
    \textrm{KE} \equiv \frac{1}{2} \rho_0 \mathbf{u}^2 \\
    \textrm{APE} = \overbrace{- \int_0^\eta g \xi \partial \rho_\textrm{nm}(z- \xi) d \xi}^{\textrm{APE}_\textrm{e}} + \int_{z-\eta}^z g \rho_\textrm{nm}(\xi) d \xi  \\
    \textrm{APV} \equiv
(\nabla \times {\bf u})\cdot \hat{\bf z} - f_0 \partial_z \eta - (\nabla \times {\bf u}) \cdot \nabla \eta
\end{align}
\end{subequations}
which all vanish in the no-motion state and satisfy the conservation laws,
\begin{equation}
\label{eqn:nonlinear-conservation-laws}
    \ddt  \left( \textrm{KE} + \textrm{APE} \right) = -\mathbf{u} \cdot \nabla  p_\textrm{tot} \quad \textrm{and} \quad  \ddt  ( \textrm{APV} ) = 0.
\end{equation}
The total energy and two materially conserved quantities lead to three global conservation requirements,
\begin{subequations}
\begin{align}
    \partial_t \int \left( \textrm{KE} +  \textrm{APE}_\textrm{e} \right) dV = 0, \\
     \int \textrm{APV} \, dV = 0, \\
      \textrm{and} \quad \int \eta \, dV = 0.
\end{align}
\end{subequations}
The thermodynamic equation \eqref{thermodynamic-eta} could be replaced with $\ddt \textrm{MCPE}=0$ from \eqref{eqn:mcpe} which would more directly lead to APE, but \eqref{thermodynamic-eta} is arguably more useful as an evolution equation. The definition of APE in \eqref{eqn:ape-apv-recap} uses the integral form of $\textrm{APE}_\textrm{e}$ from \cite{holliday1981-jfm} which results in an expression that, although less compact than $g \eta \rho_\textrm{nm}(z - \eta)$, separates the positive definite $\textrm{APE}_\textrm{e}$ from the hydrostatic pressure work.  Recall that the $\textrm{APE}_e$ is the contribution available for conversion to kinetic energy.

Linearizing \eqref{x-momentum-eta}-\eqref{continuity-eta} requires discarding any quadratic combinations of $(u,v,w,\eta)$, but also expanding $\rho_\textrm{nm}(z-\eta)$ using the lowest order approximation to $\eta$, which we write explicitly in terms of $\eta_\textrm{e}=-\rho_\textrm{e} \left( \partial_z \rho_\textrm{nm}\right)^{-1}$. As a result we have that
\begin{subequations}
\begin{align}
\label{x-momentum-eta-lin}
\partial_t u - f_0 v &= - \frac{1}{\rho_0} \partial_x p_\textrm{tot}\\ \label{y-momentum-eta-lin}
\partial_t v + f_0 u &= - \frac{1}{\rho_0}  \partial_y p_\textrm{tot}  \\ \label{z-momentum-eta-lin}
\partial_t w  &= - \frac{1}{\rho_0}  \partial_z p_\textrm{tot} - \frac{1}{\rho_0}  g \rho_\textrm{nm}(z) - N^2(z) \eta_\textrm{e} \\
% \partial_t w  &= - \frac{1}{\rho_0} \partial_z p_\textrm{tot} - \frac{g}{\rho_0} \left[\rho_\textrm{nm}(z) - \partial_z \rho_\textrm{nm}(z) \eta_\textrm{e} \right] \\ \label{thermodynamic-eta-lin}
\partial_t \eta_\textrm{e}  &= w  \\ \label{continuity-eta-lin}
\partial_x u + \partial_y v + \partial_z w &= 0, 
\end{align}
\label{eqn:boussinesq-eta-lin}
\end{subequations}
where it is understood that $\left(u,v,w,p_\textrm{tot}\right)$ are also lower order approximations of the nonlinear variables from \eqref{x-momentum-eta}-\eqref{continuity-eta}. Unlike the more traditional formulations of PE and PV, expansions of the APE and APV \eqref{eqn:ape-apv-recap} lead directly to the conserved quantities for the linearized equations of motion
\begin{subequations}
\label{eqn:linear-ke-ape-apv}
\begin{align}
    \textrm{KE} = \frac{1}{2} \rho_0 \mathbf{u}^2, \\ 
    \textrm{APE} \approx \underbrace{\frac{1}{2} \rho_0 \eta_\textrm{e}^2 N^2}_{\textrm{APE}_\textrm{e}} + g \eta_\textrm{e} \rho_\textrm{nm}(z), \\
    \textrm{APV} \approx \textrm{QGPV} = 
\partial_x v - \partial_y u - f_0 \partial_z \eta_\textrm{e}
\end{align}
\end{subequations}
after discarding the cubic terms in APE and quadratic terms in APV. The conservation laws
\begin{equation}
\label{eqn:linear-conservation-laws}
    \partial_t \left( \textrm{KE} + \textrm{APE} \right) = -\mathbf{u} \cdot \nabla  p_\textrm{tot} \quad \textrm{and} \quad \partial_t( \textrm{QGPV} ) = 0
\end{equation}
follow from the linear equations of motion \eqref{x-momentum-eta-lin}-\eqref{continuity-eta-lin}. The total energy and two materially conserved quantities lead to three global conservation requirements,
\begin{subequations}
\begin{align}
    \partial_t \int \left( \textrm{KE} +  \textrm{APE}_\textrm{e} \right) dV = 0\\
    \quad \int \textrm{QGPV} \, dV = 0 \\
    \textrm{and} \quad \int N^2 \eta_\textrm{e} \, dV = 0.
    \end{align}
\end{subequations}
The hydrostatic pressure work term $g \eta_\textrm{e} \rho_\textrm{nm}(z)$ disappears with the volume integral of APE, which can be seen by including it in the definition of pressure, or equivalently considering the perturbation equations with excess pressure and density. Noteworthy here too is that the linearization of $\int \eta dV = 0$ is derived by approximating density conservation,
\begin{equation}
    \ddt \rho_\textrm{nm}(z-\eta) = 0 \implies \ddt \left( \rho_\textrm{nm}(z) - \partial_z \rho_\textrm{nm} \eta_\textrm{e} + O\left( \eta_\textrm{e}^2 \right) \right) = 0
\end{equation}
which, after removing no-motion state,
results in $\int N^2 \eta_\textrm{e} dV = 0$. This is also exactly the nonlinear global conservation law for excess density \eqref{excess-volume-conservation}.

Finally, we note that the conservation laws for linear density ($\int N^2 \eta_\textrm{e} dV = 0)$ and PV ($\int \text{QGPV} dV =0$) %conservation laws 
impose boundary conditions on linear geostrophic solutions with zero horizontal wavenumber which are important for defining a complete basis. At other wavenumbers one can adopt the same boundary conditions or those described in \cite{smith2013-jpo} and orthogonality of the linear geostrophic solutions is ensured. Under this setting, linear geostrophic solutions can carry buoyancy anomalies at the surface. More details will be provided in an upcoming paper where we also treat the free-surface case.

%  In any case, the derived boundary conditions ensure orthogonality of linear geostrophic solutions. 

%%%%%%%%%%%%%%%%%%%%%%%%
%
\section{Concluding Remarks}
\label{sec:conclusions}
%
%%%%%%%%%%%%%%%%%%%%%%%%

While both total density ($\rho_\textrm{tot}$) and isopycnal deviation ($\eta$) lead to equally valid formulations of the equations of motion, \eqref{x-momentum}-\eqref{continuity} and \eqref{x-momentum-eta}-\eqref{continuity-eta}, respectively, the $\eta$-formulation has the significant advantage that the potential energy and potential vorticity are naturally expressed as APE and APV in those variables. The APE follows directly from the materially conserved potential energy \eqref{eqn:mcpe} and has two separable components: a positive definite contribution from $\textrm{APE}_\textrm{e}$ available for conversion to kinetic energy, and a contribution from work against the hydrostatic pressure force. This framework clarifies that the transition from the full equations to the perturbation equations also implies removal of the contributions of hydrostatic pressure work on each fluid parcel, resulting in the definition of $\textrm{APE}_\textrm{e}$ as in \cite{holliday1981-jfm}. The framework 
%here 
is consistent with ideas laid out in \cite{winters1995-jfm} and thus can 
%also 
be used as a tool to study diapycnal processes that alter the APE.

The APV results from Ertel PV defined using $z-\eta$, and leads to an exact, analytical expression for potential vorticity that vanishes in the limit of no-motion. 
 Importantly, the APV also results in the QGPV in its linear limit, consistent with the quantity derived from the linearized equations of motion. This is important because the quality of approximation can be directly assessed by comparing QGPV with the full APV. Furthermore, boundary conditions for the linear modes that ensure orthogonality can be obtained from physical principles, namely density conservation and conservation of QGPV. Our goal is to extend the analysis to flows with a free surface and this paper is one step in that direction.

Declaration of Interests. The authors report no conflict of interest.

\bibliographystyle{jfm}

\end{document}